\documentclass[preprint,12pt]{elsarticle}
\usepackage{amsmath}
\usepackage{amssymb}
\usepackage[dvipdfm,colorlinks]{hyperref}
\usepackage{multirow}

\biboptions{compress}

\begin{document}
\begin{frontmatter}


\title{Quantum McEliece public-key encryption scheme}
\author[label1]{Li Yang}\ead{yangli@iie.ac.cn}
\author[label2]{Min Liang}\ead{liangmin07@mails.ucas.ac.cn}
\address[label1]{State Key Laboratory of Information Security, Institute of Information Engineering, Chinese Academy of Sciences, Beijing 100093, China}
\address[label2]{Data Communication Science and Technology Research Institute, Beijing 100191, China}

\begin{abstract}
This paper investigates a quantum version of McEliece public-key encryption (PKE) scheme, and analyzes its security. As is well known, the security of classical McEliece PKE is not stronger than the onewayness of related classical one-way function. We prove the security of quantum McEliece PKE ranks between them. Moreover, we propose the double-encryption technique to improve its security, and the security of the improved scheme is proved to be between the original scheme and the quantum one-time pad.
\end{abstract}


\end{frontmatter}
\journal{}


\section{Introduction}
Public-key encryption (PKE) is one of the most important research directions in modern cryptography, and has been widespread used in information communication. However, the widely used PKE schemes, such as RSA have been threatened by quantum attack. Then it becomes important to construct PKE scheme against quantum attack.

Okamoto et al.~\cite{Okamoto00} constructed the first quantum PKE scheme based on subset-sum problem, whose public-key is computed from the private-key with Shor's algorithm for finding discrete logarithm, though the private-key, public-key, plaintext and ciphertext are all classical. In~\cite{Kawachi05}, a quantum PKE is constructed based on a hard problem $QSCD_{ff}$, which has been proved to be one with bounded information theoretic security \cite{hayashi08}. By using single-qubit rotations, Nikolopoulos~\cite{Nikolopoulos08} proposed a quantum PKE with classical private-key and quantum public-key. Based on quantum encryption, Gao et al.~\cite{Gao09} presented a quantum PKE with symmetric keys, with two qubits from a Bell state serving as the public-key and the private-key, respectively. Pan and Yang~\cite{Pan10} constructed a quantum PKE scheme with information theoretic security. All these quantum PKEs above are classical bits oriented.
However, quantum messages also need to be encrypted in some cases. Though quantum key distribution (QKD) plus quantum one-time pad (QOTP) can finish the task of encrypting quantum messages, it needs some preshared keys in the implementation of QKD. This paper explores the asymmetric scheme of this task, and propose a quantum-message-oriented PKE.

McEliece public-key encryption scheme \cite{McEliece78} is based on coding theory and its security relies on the difficulty of solving a NP-complete (NPC) problem. Though the scheme is a classical PKE scheme, it is believed that it can resist quantum attack. Based on its construction, the researchers begin to construct PKE scheme in quantum world, for the purpose of encryption of quantum messages. Yang \cite{Yang05} proposed the first quantum analogue of McEliece PKE, in which the public-key and private-key are classical, however it can encrypt quantum messages. Later in 2010, we extend it and present the definition of induced trapdoor one-way transformation (OWT), then construct a framework of quantum PKE based on the induced trapdoor OWT. The quantum McEliece PKE proposed in \cite{Yang05} can be seen as a special case of the quantum PKE framework. In 2012, Fujita \cite{fujita2012} also proposed a quantum analogue of McEliece PKE based on quantum coding theory, and its security also relies on the difficulty of solving a NPC problem. This scheme also uses the classical keys and can encrypt quantum messages.

This paper studies the security of the quantum McEliece PKE scheme which is proposed in Ref.\cite{Yang05,yang2010}, and then focuses on the improvement to it.

\section{Quantum public-key encryption}
Firstly, we define quantum public-key encryption(QPKE) as follows. Without loss of generality, the definition is presented for the encryption of quantum messages (The classical messages can be seen as a special case).

{\bf Definition 1:} A quantum public-key encryption scheme is described by a triplet $(\mathcal{G},\mathcal{E},\mathcal{D})$, where
\begin{enumerate}
  \item $\mathcal{G}$ is the polynomial time quantum key-generation algorithm. On input $1^n$, $\mathcal{G}$ outputs $(e, d)$ in polynomial time, where $e$ is a public key, $d$ is a secret-key, and $n$ is a security parameter.
  \item $\mathcal{E},\mathcal{D}$ are the polynomial time quantum encryption/decryption algorithms. They satisfy this condition:
  For every $n$-qubit message $\sigma$, every polynomial $poly(n)$, and all sufficiently large $n$, $$F(\mathcal{D}(\mathcal{E}(\sigma, e), d),\sigma)>1-1/poly(n),$$ where $F(\sigma_1,\sigma_2)$ denotes the fidelity of two states $\sigma_1,\sigma_2$.
\end{enumerate}

Next, we present the security definition of QPKE.

{\bf Definition 2:} A quantum public-key encryption scheme is computationally (information-theoretically) secure, if for every polynomial-size (unlimited-size) quantum circuit family ${C_n}$, every positive polynomial $p(.)$, all sufficiently large $n$, and any two quantum messages $\sigma,\sigma'\in H_M$, it holds that
$$\left|Pr[C_n(\mathcal{G}(1^n),\mathcal{E}_{\mathcal{G}(1^n)}(\sigma))=1]-Pr[C_n(\mathcal{G}(1^n),\mathcal{E}_{\mathcal{G}(1^n)}(\sigma'))=1]\right| <\frac{1}{p(n)},$$
where $\mathcal{E}$ is a polynomial time quantum encryption algorithm and $\mathcal{G}$ is a polynomial time quantum algorithm for generating public-keys.

\section{\label{seccount2}Quantum McEliece public-key encryption scheme}
\subsection{Some notations}
Suppose $\rho=\sum_{m\in\{0,1\}^k}\sum_{m'\in\{0,1\}^k}\alpha_{mm'}|m\rangle\langle m'|$ and $G$ is a $k\times n$ matrix, then we denote $$\rho\circ G=\sum_m\sum_{m'}\alpha_{mm'}|mG\rangle\langle m'G|,$$ where $mG$ is the multiplication of the vector $m$ and matrix $G$ modular 2.

Suppose $x$ is arbitrary vector in $\{0,1\}^k$, then we denote $$\rho\dotplus x=\sum_{m\in\{0,1\}^k}\sum_{m'\in\{0,1\}^k}\alpha_{mm'}|m+x\rangle\langle m'+x|,$$ where $m+x$ is the bitwise addition of $m$ and $x$ modular 2.

Suppose a matrix $M$ is a $n\times n$ invertible matrix, then denote $M^{-1}$ as the inverse matrix of $M$.

Suppose a matrix $M$ is a $k\times n$ ($k<n$) matrix and it is full row rank, then it has Moore-Penrose inverse. Denote $M^{-}$ as one of Moore-Penrose inverses of $M$ satisfying $MM^{-}=I$ ($I$ is identity matrix).

\subsection{\label{seccount5}Scheme~\textrm{\cite{Yang05}}}
Quantum McEliece public key encryption scheme is firstly proposed in Ref.\cite{Yang05}. This scheme will be briefly introduced before our analysis and improvement.

The quantum key-generation algorithm is the same as classical McEliece PKE protocol~\cite{McEliece78}: Suppose $G$ is a $k\times n$ generator matrix of a $[n,k,d]$ Goppa code,
$G'=SGP$, here $S$ is a $k\times k$ invertible matrix and $P$ is an $n\times n$ permutation matrix.
We choose $(G',t),t\leq\lfloor\frac{d-1}{2}\rfloor$ as the public-key and $(S,G, P)$ as the private-key. Let $H$ is the check matrix of Goppa code
satisfying $GH^T=0$.

Alice selects a random number $r$ of weight $\leq t$, and uses Bob's public-key $G'$ with $r$ to encrypt a $k$-qubit state $\rho$. This encryption can be shown by the density as follows.
$$\rho\rightarrow\rho\circ G'\rightarrow\rho\circ G'\dotplus r.$$
Denote $\rho\circ G'\dotplus r\triangleq \rho_c$, the $\rho_c$ is the quantum ciphertext.
The above transformation is feasible. The reason will be shown later.

Bob uses his private-key $s=(S,G,P)$ to decrypt the state $\rho_c$ coming from Alice:
Firstly he computes the state $\rho_c\circ P^{-1}$(=$\rho\circ SG\dotplus rP^{-1}$) and extract the value of $rP^{-1}$; Then he computes $\rho_c\circ P^{-1}\dotplus rP^{-1}$(=$\rho\circ SG$), and can further obtain the state $$((\rho_c\circ P^{-1}\dotplus rP^{-1})\circ G^{-})\circ S^{-1},$$
which is equal to $((\rho\circ SG)\circ G^-)\circ S^{-1}\rho=(\rho\circ S)\circ S^{-1}=\rho$. Note that, $G^-$ is a Moore-Penrose inverse of $G$. So $G^-$ is a $n\times k (n>k)$ binary matrix.
According to Proposition 5 in the appendix, the transformation $\sigma\circ G^-$  is infeasible physically for arbitrary $n$-qubit state $\sigma$. However, $\rho\circ SG$ is a special subclass of all $n$-qubit states, which is related to $G$. So the computation $(\rho\circ SG)\circ G^-$ is feasible physically. This will be shown in the following concrete scheme, see Eq.(\ref{equ1}).

Next, we show the encryption/decryption algorithms in the Dirac form, which is a more understandable way.

Denote the $k$-qubit state $\rho$ as $\sum_m\alpha_m|m\rangle$. Then, this encryption can be described in the following three steps
\begin{eqnarray}
|r\rangle\sum_m\alpha_m|m\rangle|0\rangle
&\rightarrow& |r\rangle\sum_m\alpha_m|m\rangle|mG'\rangle
\rightarrow |r\rangle\sum_m\alpha_m|m\oplus mG'G'^-\rangle|mG'\rangle \nonumber\\
&\rightarrow& |r\rangle|0\rangle\sum_m\alpha_m|mG'\oplus r\rangle,
\end{eqnarray}
where the matrix $G'^-$ is a generalized inverse matrix of $G'$. Because $G'$ is a full row
rank matrix, there exists $G'^-$ that satisfies $G'G'^-=I_k$. This is the condition that one can get $\sum_m\alpha_m|mG'\rangle$ from $\sum_m\alpha_m|m\rangle$. Alice sends the cipher state $\sum_m\alpha_m|mG'\oplus r\rangle$ to Bob.

Bob uses his private-key $s=(S,G,P)$ to decrypt the state coming from Alice,
\begin{eqnarray}
&&|s\rangle\sum_m\alpha_m|mG'\oplus r\rangle|0\rangle|0\rangle \rightarrow |s\rangle\sum_m\alpha_m|mG'\oplus r\rangle|(mG'\oplus r)P^{-1}\rangle|0\rangle \nonumber\\
&\rightarrow&|s\rangle\sum_m\alpha_m|0\rangle|(mG'\oplus r)P^{-1}\rangle|0\rangle=|s\rangle|0\rangle\sum_m\alpha_m|mSG\oplus rP^{-1}\rangle|0\rangle \nonumber\\
&\rightarrow&|s\rangle|0\rangle\sum_m\alpha_m|mSG\oplus rP^{-1}\rangle|(mSG\oplus rP^{-1})H^T\rangle \nonumber\\
&&=|s\rangle|0\rangle\sum_m\alpha_m|mSG\oplus rP^{-1}\rangle|rP^{-1}H^T\rangle,
\end{eqnarray}
then measures the second register to get $rP^{-1}H^T$, and find $rP^{-1}$ via the fast decoding
algorithm of the Goppa code generated by $G$. Bob carries out the following transformation on the quantum state $\sum_m\alpha_m|mSG\oplus rP^{-1}\rangle$ according to the value of $rP^{-1}$,
\begin{equation}
|rP^{-1}\rangle\sum_m\alpha_m|mSG\oplus rP^{-1}\rangle\rightarrow|rP^{-1}\rangle\sum_m\alpha_m|mSG\rangle.
\end{equation}
Then he computes
\begin{eqnarray}\label{equ1}
&&|s\rangle\sum_m\alpha_m|mSG\rangle|0\rangle|0\rangle \nonumber\\
&\rightarrow&|s\rangle\sum_m\alpha_m|mSG\rangle|mSGG^-\rangle|0\rangle=|s\rangle\sum_m\alpha_m|mSG\rangle|mS\rangle|0\rangle \nonumber\\
&\rightarrow&|s\rangle\sum_m\alpha_m|0\rangle|mS\rangle|0\rangle \nonumber\\
&\rightarrow&|s\rangle|0\rangle\sum_m\alpha_m|mS\rangle|mSS^{-1}\rangle=|s\rangle|0\rangle\sum_m\alpha_m|mS\rangle|m\rangle \nonumber\\
&\rightarrow&|s\rangle|0\rangle|0\rangle\sum_m\alpha_m|m\rangle.
\end{eqnarray}
Finally, the quantum message $\sum_m\alpha_m|m\rangle$ is obtained.

\subsection{Analysis}\label{sec22}
Induced trapdoor one-way transformation (OWT) has been defined in Ref. \cite{yang2010}. The above protocol satisfies the framework of QPKE based on induced trapdoor OWT.
Let $g(m,r)=0$ and $f(m,r)=mG'\oplus r$ in the induced trapdoor OWT.
Here $g(m,r)$ is a constant function, then the encryption transformation can be simplified as
$$U_{fg}(r)=\sum_m|0\rangle\langle m|\otimes|mG'\oplus r\rangle\langle 0|.$$
The decryption transformation is
$$D_{fg}(s)=\sum_{r,m}|r(f,s)\rangle\langle 0|\otimes|m\rangle\langle 0|\otimes|0\rangle\langle f(m,r)|.$$
where $r(f,s)$ denotes a function that is relative to $f,s$.

\begin{itemize}
\item{\it Firstly, we analyze its security while encrypting classical messages.}
\end{itemize}

The quantum McEliece PKE scheme can be used to encrypt classical message. In this case, the quantum McEliece PKE would degenerate to the corresponding classical PKE. The classical McEliece PKE has been studied for more than thirty years in modern cryptography, and is believed to be secure. Thus, our scheme is secure while encrypting classical messages.

Ref.\cite{fujita2012} believes our scheme is insecure when encrypting classical messages. Though the McEliece PKE has not been reduced to NP-complete problem, here we discuss the difficult from a new view when attacking the ciphertext $mG'\oplus r$.
Attacking cipher $mG'\oplus r$ is equivalent to attacking $m\oplus rG'^{-}$.
Now,we show the difficulty of decoding $m\oplus rG'^{-}$.

As it is in McEliece PKE scheme,
we know $G'=SGP$, where $S,P$ are both invertible matrices, $G$ is generator matrix of Goppa code and is full row rank, so $G'$ is also full row rank, and then it has Moore-Penrose inverse. Suppose $G_1^{'-}$ is one of Moore-Penrose inverses of $G'$ satisfying $G'G_1^{'-}=I$. In fact, $G_1^{'-}$ can be obtained by solving the linear equations $G'X=I$. Then all the Moore-Penrose inverses of $G'$ can be written as the form
$$G'^-=G_1^{'-}\oplus U\oplus G_1^{'-}G'U,$$
where $U$ is any $n\times k$ binary matrix. It can be verified that $G'G'^-=I$.
In classical McEliece PKE scheme, the cipher $c$ and plaintext $m$ satisfy the relation $c=mG'\oplus r$, where $r$ is a binary row vector of weight $t$.
Suppose Eve finds another Moore-Penrose inverse of $G'$, denoted as $G_2^{'-}$, then he can compute $cG_2^{'-}=m\oplus rG_2^{'-}$.
Denote $G_2^{'-}=(e_1\cdots e_k)$, where each $e_i$ is a binary column vector. Then $cG_2^{'-}$ can be represented as $(m_1\oplus r\cdot e_1),\cdots,(m_k\oplus r\cdot e_k)$.
If each column $e_i$ of $G_2^{'-}$ has more zeros (it means the Hamming weight of $e_i$ is small enough), $r\cdot e_i$ would equal to $0$ with large probability, then its $i-$th bit $m_i\oplus r\cdot e_i$ would reveal the $i-$th bit of original plaintext with large probability.
Notice that $e_i=g_i\oplus (I\oplus G_1^{'-}G')u_i$, where $g_i$ and $u_i$ are the $i-$th column of $G_1^{'-}$ and $U$ separately.
Here $g_i$ and $I\oplus G_1^{'-}G$ are known, but $u_i$ is unknown. Now Eve have to face a problem: finding $u_i$,
such that $g_i\oplus (I\oplus G_1^{'-}G')u_i$ has weight smaller than a given value. This is just a LPN problem,
which is a NP-complete problem.

{\it Remark 1:} This problem can be seen from another view. $I\oplus G_1^{'-}G'$ is a $n\times n$ matrix, and $g_i,u_i$ are two $n\times 1$ vectors, but $u_i$ is unknown. So the above problem can be restated as follow: how to select some columns of $I\oplus G_1^{'-}G'$, such that their summation is closest to vector $g_i$? This is just a closest vector problem (CVP), which is a NP-hard problem.

We have tried numerical experiment following the above attack, and it seems that this kind of attack is invalid.
This attack is reduced to an optimal problem (CVP): finding $u_i$, such as $g_i\oplus (I\oplus G_1^{'-}G')u_i$ (notice that it equals $e_i$) has weight smaller than a given value.
Suppose the parameters $n=1024,k=524,t=50$ in the McEliece PKC scheme.
Firstly, because $u_i$ has $2^{1024}$ choices, both the exhaustive search and random search are not realistic. While choosing some small parameters such as $n=60,k=30$, the exhaustive search can reduce the weight of $e_i$ to $1$ with probability $2\%$, and the random search may be slightly better.  With the greedy search, we obtain $e_i$ of weight $225$ on average. In this case, $Pr[r\cdot e_i=0]\approx 0.5+0.1\times 10^{-13}$, here $r$ is a $n$-bit random vector of weight $t=50$. Thus, the attack presented here is invalid.

\begin{itemize}
\item{\it Secondly, we strictly prove the relationship between the security of quantum PKE protocols and that of its classical counterpart.}
\end{itemize}

{\bf Theorem 1:} The quantum McEliece PKE is at least as secure as classical McEliece PKE protocol.

{\bf Proof:}
Suppose there is a quantum algorithm $A$, which can efficiently transform the
cipher state $\sum_m\alpha_m|mG'\oplus r\rangle$ into quantum message $\sum_m\alpha_m|m\rangle$.
In order to decrypt arbitrary classical cipher $m_0G'\oplus r_0$,
we firstly prepare a quantum state $|m_0G'\oplus r_0\rangle$.
Then, the quantum state $|m_0G'\oplus r_0\rangle$ is an input to the quantum algorithm $A$,
and will be transformed into the quantum state $|m_0\rangle$. Finally, the classical message $m_0$ is obtained via measuring the output quantum state $|m_0\rangle$. Thus, if there is an attack to quantum McEliece PKE, there would be an attack to classical McEliece PKE. Therefore, quantum McEliece PKE is at least as secure as classical McEliece PKE protocol.$\hfill{~}\Box$

Since the functions $f(m,r)$ and $g(m,r)$ are classical functions,
finding the trapdoor $s$ is a classical computational problem. Thus, the security of QPKE protocol
based on induced trapdoor OWT depends on the onewayness of corresponding classical trapdoor one-way function.

Now we can arrive at the following conclusion.

{\it The security of a QPKE protocol based on induced trapdoor OWQT is one
between that of corresponding classical PKE and the onewayness of related classical one-way function.
In other words, the security of a QPKE given above is not stronger than the onewayness of related classical
trapdoor one-way function, and is not weaker than the security of its classical counterparts.}


\begin{itemize}
\item{\it Finally, we analyze its security from the aspect of attack.}
\end{itemize}

The attacker can have two different strategies: 1) attacking the secret key from the public key; 2) attacking the ciphertext $\sum_m\alpha_m|mG'\oplus r\rangle$.

If the attacker adopts the first strategy, the difficulty is the same as that of attacking the classical McEliece PKE, because the quantum PKE scheme uses the same key-generating algorithm as its classical counterpart.

If the attacker adopts the second strategy, we should analyze what can be extracted from the ciphertext $\sum_m\alpha_m|mG'\oplus r\rangle$.

{\bf Theorem 2:} The strategy 2) is inefficient when attacking the ciphertext of quantum McEliece PKE scheme.

{\bf Proof:}
Because $k\times n$ matrix $G'$ is public and full rank $k$ ($k<n$), the attacker can compute its generalized inverse matrix $G'^-$, which is a $n\times k$ matrix. Thus, the attacker can perform the following processing on the ciphertext.
\begin{eqnarray}
\sum_m\alpha_m|mG'\oplus r\rangle|0\rangle
&\rightarrow& \sum_m\alpha_m|mG'\oplus r\rangle|m\oplus rG'^-\rangle \nonumber\\
&\rightarrow& |r\oplus rG'^-G'\rangle\sum_m\alpha_m|m\oplus rG'^-\rangle.
\end{eqnarray}
Then, the attacker can measurement the first register, and obtain the value of $r(I\oplus G'^-G')$. In addition, he can also obtain a new ciphertext $\sum_m\alpha_m|m\oplus rG'^-\rangle$, which can be written as $X(rG'^-)\sum_m\alpha_m|m\rangle$ or $\sum_m\alpha_m|(mG'\oplus r)G'^-\rangle$.
From the following two propositions, the proof can be finished.

{\bf Proposition 1:} Given the values of $r(I\oplus G'^-G')$ and $I\oplus G'^-G'$, solving the value of $r$ is a LPN problem.

{\bf Proof:}
Now we know the attacker can obtain the values of $r(I\oplus G'^-G')$ and $I\oplus G'^-G'$. Because $G'(I\oplus G'^-G')=G'\oplus G'=0$ and $G'\neq 0$, it can be inferred that $n\times n$ matrix $I\oplus G'^-G'$ is not full rank.
So only a little information about $r$ cannot be computed from the values of $r(I\oplus G'^-G')$ and $I\oplus G'^-G'$. However, the value of $rG'^-$ is still hard to compute. Denote $v=r(I\oplus G'^-G')$. Then, computing the value of $rG'^-$ is a NPC problem: given the values of $v$,$G'$, $r$ is a random binary vector, how to compute the value of $rG'^-$ from the equation $v=r\oplus (rG'^-)G'$? It is just a LPN problem which has been discussed above. $\hfill{~}\Box$

Now we know the attacker can transform the ciphertext $\sum_m\alpha_m|mG'\oplus r\rangle$ into $\sum_m\alpha_m|(mG'\oplus r)G'^-\rangle$. However, it does not hold for any quantum state $\sum_m\alpha_m|m\rangle$. In other words, the transformation $\sum_m\alpha_m|m\rangle\rightarrow\sum_m\alpha_m|mH\rangle$ may be physically infeasible. The proof is given in the appendix.

{\bf Proposition 2:} The state $X(rG'^-)\sum_m\alpha_m|m\rangle$ is unrelated with the plaintext $\sum_m\alpha_m|m\rangle$, from the view of fidelity.

{\bf Proof:}
According to Fujita's analysis \cite{fujita2012}, $X$ basis measurement on the two states $X(rG'^-)\sum\nolimits_m\alpha_m|m\rangle$ and $\sum\nolimits_m\alpha_m|m\rangle$ can result a same statistical probability.
This is obviously correct since the two states differ only in some bit-flips \cite{Nielsen00}.
So it is expected that the attacker may obtain some information about the quantum messages by quantum measurement on $X(rG'^-)\sum\nolimits_m\alpha_m|m\rangle$.

Though there exists a vulnerability in quantum McEliece PKE, it should be stressed that similarity of these two states $X(rG'^-)\sum_m\alpha_m|m\rangle$ and $\sum_m\alpha_m|m\rangle$ is described by the fidelity of them:
\begin{eqnarray}
F(e) &=& \left|(\sum_m\alpha_m^*\langle m|)\left|X(e)\right|(\sum_n\alpha_n|n\rangle)\right| \nonumber\\
&=& \left|\sum_{m,n}\alpha_m^*\alpha_n\langle m|X(e)|n\rangle\right|=\left|\sum_m\alpha_m^*\alpha_{m\oplus e}\right|,
\end{eqnarray}
where $e=rG'^-$ is a random string depending on the error $r$. It can be seen that $F(e)$ may equals to any value from 0 to 1, then, generally speaking, identical probability distributions do not means identical states.$\hfill{~}\Box$

{\it Remark 2:} The attack to the state $X(rG'^-)\sum_m\alpha_m|m\rangle$ can also be analyzed from information theory. According to Holevo theorem \cite{Nielsen00}, the quantum measurement on
$X(r_1G_1'^-)\sum_m\alpha_m|m\rangle$ can obtain at most $k$-bit information, but $\sum_m\alpha_m|m\rangle$ has $2^k$ amplitudes $\alpha_m$. Suppose each amplitude is accurate to $l$ decimal places, then each $\alpha_m$ can be seen as $l$-bit complex number which has both real and image parts, so it is necessary to obtain $2l\times2^{k}$-bit information for determining an unknown state $\sum_m\alpha_m|m\rangle$. It can be seen that even Alice encrypts the same quantum state polynomial times, the attacker can obtain at most a polynomial-bits information. It is still hard for her to determine the state $\sum_m\alpha_m|m\rangle$.

\section{Double-encryption scheme}
\subsection{Scheme}
As stated by Fujita \cite{fujita2012}, the vulnerability of quantum McEliece PKE is due to the fact that our PKC introduces no phase encryption. In this section, we propose an improved variant of the quantum McEliece PKE scheme using double-encryption technique.

The encryption is briefly stated as follows. The quantum McEliece PKE is used twice, however, the second encryption uses the different parameters from the first.
\begin{enumerate}
\item Alice uses two pairs of public-keys $(G_1',t_1)$ and $(G_2',t_2)$. She firstly uses the first public-key $(G_1',t_1)$ to encrypt the $k$-qubit message $\sum_m\alpha_m|m\rangle$, and obtain a $n$-qubit state $\sum_m\alpha_m|mG_1'\oplus r_1\rangle$.
\item Then she performs an Hadamard transformation $H^{\otimes n}$ on this state and obtain $H^{\otimes n}\sum_m\alpha_m|mG_1'\oplus r_1\rangle=\sum_m\alpha_m\sum_k(-1)^{k\cdot(mG_1'\oplus r_1)}|k\rangle$.
\item Finally she uses the second public-key $(G_2',t_2)$ to encrypt the $n$-qubit quantum state $H^{\otimes n}\sum_m\alpha_m|mG_1'\oplus r_1\rangle$, and obtains a $n'$-qubit quantum state, which is the ciphertext of the improved scheme. The final ciphertext is as follow:
    \begin{eqnarray}
    &&\sum_m\alpha_m\sum_k(-1)^{k\cdot(mG_1'\oplus r_1)}|kG_2'\oplus r_2\rangle \nonumber\\
    =&&\sum_k\left[\sum_m\alpha_m(-1)^{k\cdot(mG_1'\oplus r_1)}\right]|kG_2'\oplus r_2\rangle.
    \end{eqnarray}
\end{enumerate}

Bob receives the ciphertext, and performs the following decryption process.
\begin{enumerate}
\item Bob uses the second private-key to decrypt the received $n'$-qubit quantum cipher state, and obtain a $n$-qubit state $\sum_m\alpha_m\sum_k(-1)^{k\cdot(mG_1'\oplus r_1)}|k\rangle$.
\item Then he performs an Hadamard transformation $H^{\otimes n}$ on the $n$-qubit state.
\item Finally he uses the first private-key to decrypt and obtain the $k$-qubit message.
\end{enumerate}

\subsection{Analysis}
Firstly, it should be noticed that, our scheme is more simple than the scheme proposed by Fujita \cite{fujita2012}. Fujita's scheme is constructed based on quantum error-correction code, which has the ability to correct quantum errors. However, the encoding in our scheme uses classical error-correction code and cannot correct quantum errors, and has less redundance, so its encoding circuit needs less ancillary qubits and is more simple than Fujita's scheme.

Now, let's consider the security of double-encryption scheme. From the view of Alice, the attacker can obtain the following quantum state by performing a unitary about $G_2'^-$,
$$X(r_2G_2'^-)\sum_k\left[\sum_m\alpha_m(-1)^{k\cdot(mG_1'\oplus r_1)}\right]|k\rangle.$$
Then he performs $H^{\otimes n}$ and obtains the following state
\begin{eqnarray}
&&H^{\otimes n}X(r_2G_2'^-)\sum_k\left[\sum_m\alpha_m(-1)^{k\cdot(mG_1'\oplus r_1)}\right]|k\rangle \nonumber\\
&=& Z(r_2G_2'^-)H^{\otimes n}\sum_k\left[\sum_m\alpha_m(-1)^{k\cdot(mG_1'\oplus r_1)}\right]|k\rangle \nonumber\\
&=& Z(r_2G_2'^-)\sum_m\alpha_m|mG_1'\oplus r_1\rangle \nonumber\\
&=& \sum_m\alpha_m(-1)^{(r_2G_2'^-)\cdot(mG_1'\oplus r_1)}|mG_1'\oplus r_1\rangle,
\end{eqnarray}
and then perform a transformation with relative to $G_1'^-$, and finally obtains a state \begin{equation}\label{equ6}X(r_1G_1'^-)\sum_m\alpha_m(-1)^{(r_2G_2'^-)\cdot(mG_1'\oplus r_1)}|m\rangle.\end{equation}
During the above process, the attacker can obtain the values of $r_2(I\oplus G_2'^-G_2')$,$r_1(I\oplus G_1'^-G_1')$, $I\oplus G_2'^-G_2'$ and $I\oplus G_1'^-G_1'$. However, he still cannot obtain the values of $r_1,r_2$. The reason is the same as the analysis in Section \ref{sec22}.

Then, whether one can extract some information about the plaintext $\sum_m\alpha_m|m\rangle$ from the state in Eq.(\ref{equ6})?


In the original quantum McEliece PKE scheme, the attacker can transform the ciphertext and obtain the state $X(rG'^-)\sum_m\alpha_m|m\rangle$. By comparing this state with the original quantum message $\sum_m\alpha_m|m\rangle$, they differs only some bit-flip errors. Thus, the original encryption scheme introduces only bit-flip errors, however, bit-flip errors can be seen as phase errors in conjugate space because of $HXH=Z$. This is so called vulnerability discussed in \cite{fujita2012}.

When it is modified with double-encryption scheme, the attacker can obtain the quantum state expressed in Eq.(\ref{equ6}).
By comparing this state with quantum message $\sum_m\alpha_m|m\rangle$, both bit-flip errors and phase errors are introduced. Thus, whether it is seen from the conjugate space or not, the two types of errors exist simultaneously. Then the vulnerability is eliminated. The detail arguments are as follows.

{\bf Theorem 3:} The double-encryption scheme is more secure than the original quantum PKE scheme in Sec.\ref{seccount2}.

{\bf Proof:}
Because the attacker can transfer the quantum cipher into the state $Z(r_2G_2'^-)\sum_m\alpha_m|mG_1'\oplus r_1\rangle$ in Eq.(\ref{equ6}), with regard to the attacker, the encryption operator can be written as
\begin{eqnarray}\label{equ9}
U(r_1,r_2) &=& Z(r_2G_2'^-)\sum_m|mG_1'\oplus r_1\rangle\langle m,0\cdots 0| \nonumber\\
&=& Z(r_2G_2'^-)X(r_1)\sum_m|mG_1'\rangle\langle m,0\cdots 0| \nonumber\\
&=& Z(r_2G_2'^-)X(r_1)V,
\end{eqnarray}
where the operator $V=\sum_m|mG_1'\rangle\langle m0|$ is independent of the two error vectors $r_1,r_2$. Given the public key $G_1'$, the operator $V$ is a constant operator.

{\it Remark 3:} As is known from Refs.\cite{ambainis2000,boykin2000}, the encryption operator in private quantum channel can be written as $U(a,b)=Z(b)X(a)$, where $a,b$ are chosen randomly, and its security depends on the randomness of $a,b$. Because $V$ in Eq.(\ref{equ9}) is a constant operator, the encryption operator $U(r_1,r_2)$ can be seen as a special kind of private quantum channel, where the difference lies in that the weight of random vectors $r_1,r_2$ is bounded by $t_1,t_2$ separately.

In the original quantum McEliece PKE scheme, the encryption operator can be seen as $U'(r_1,r_2)=Z(r_2)X(r_1G_1'^-)$, where $r_2\equiv 0$.
The random bits of $r_1,r_2$ in the double-encryption scheme is twice more than that in the original quantum McEliece PKE scheme. In other words, attacking the state $U(r_1,r_2)\sum_m\alpha_m|m\rangle$ is more difficult than attacking the state $U'(r_1,0)\sum_m\alpha_m|m\rangle$.
Thus, the double-encryption scheme can improve the security of our quantum PKE scheme.$\hfill{~}\Box$

From the above proof, one can informally conclude that

{\it To achieve the same security as the original quantum McEliece PKE scheme in Sec.\ref{seccount2}, the double-encryption scheme requires about half of the key-length than the original scheme.}

{\bf Proposition 3:} Multiple use of the double-encryption scheme will decrease its security.

{\bf Proof:}
In the double-encryption scheme (usually let the parameters $k=524$, $n=1024$, $n'=2n=2048$), the length of ciphertext is expanded about $4$ times ($n'/k\approx 4$), however, the bits of random key are expanded only $3$ times ($(n'+n)/n=3$). So, the ratio between the bit-length of the random key and the length of ciphertext would decrease approaching zero when the double-encryption scheme are used several times. That is, with respect to the length of ciphertext, the amount of key is reduced, then it means the security will be worse. Thus, the security will decrease when using multiple times of the double-encryption scheme.$\hfill{~}\Box$

{\it Remark 4:} Though the double-encryption scheme is more secure than the once-encryption scheme, it is enough to adopt once-encryption scheme (the scheme is given in Sec.\ref{seccount2}) in some low level security scenario.

Finally, it is worth to noticed that, in private quantum channel, the random numbers $r_1,r_2$ are not locally generated, and are preshared keys. However, in double-encryption scheme, $r_1,r_2$ are locally selected, and are random numbers which are used in the encryption only once. In addition, according to our scheme, two identical quantum messages may be encrypted into two different ciphertexts since the different random numbers $r_1,r_2$ are used every time. Thus, there are only one chance when attacking the ciphertext of a quantum message through quantum measurement, in other words, the message can be encrypted several times without loss of security.

\section{Discussions}


There has been several attack to classical McEliece PKE. However, an attack to classical McEliece PKE does not mean an attack to quantum McEliece PKE.
There are several kinds of attack to classical McEliece PKE, such as Korzhik-Turkin attack~\cite{Korzhik91},
message-resend attack and related-message attack~\cite{Berson97}. Since the detail of Korzhik-Turkin attack has not
been given till now, the efficiency of this attack is still an open problem. Because iterative decoding algorithm
is used in the Korzhik-Turkin attack, and quantum state cannot be reused, it fails when attacking quantum McEliece PKE.
Though classical McEliece PKE has to be improved to prevent message-resend attack
and related-message attack~\cite{Sun98}, these attacks also fail while facing the quantum McEliece PKE protocol.
Therefore, quantum McEliece PKE is more secure than classical McEliece PKE protocol.

Our quantum PKE schemes are designed to encrypt quantum message $\sum_m\alpha_m|m\rangle$.
However, if we consider the number $r$ involved as classical message encrypted, this kind of QPKE scheme
can also be regarded as `quantum envelope' for classical message transmission. In addition, since the attacks to classical McEliece PKE,
such as Korzhik-Turkin attack~\cite{Korzhik91}, message-resend attack and related-message attack~\cite{Berson97},
fail to attack quantum McEliece PKE,
it is probably more secure to transmit classical information via quantum McEliece PKE than that via classical McEliece PKE.

Actually, the quantum McEliece PKE scheme has ever been presented originally in a conference paper (see Ref.\cite{Yang05}). This paper investigates that original scheme and develops double-encryption technique to improve its security. Though we only construct the quantum version of McEliece PKE, the method here can also be extended to construct quantum versions of other classical PKE schemes. The details are presented in Ref.\cite{yang2010}. It is worth to notice that some of the schemes in Ref.\cite{yang2010} do not have post-quantum security.

The other quantum McEliece PKE proposed by Fujita \cite{fujita2012} is based on quantum coding. The security of both Fujita's and our schemes depends on the difficulty of solving NPC problem. Fujita \cite{fujita2012} pointed out a vulnerability of our scheme in Ref.\cite{Yang05}, however, it has been improved in this paper. Ref.\cite{fujita2012} argued that the PKE scheme in Ref.\cite{Yang05} is insecure while encrypting classical messages. Here we have clarified it in Sec.\ref{sec22}. In addition, we would like to mention that it is sufficient to adopt the original PKE scheme proposed in Ref.\cite{Yang05} in some low-level security scenario besides encrypting classical messages. Finally, we would argue that, our scheme is more simple than Fujita's scheme. The reason is as follows: Fujita's scheme is based on quantum error correction code and the encoding can correct quantum errors, while our scheme is based on classical error correction code and does not have the ability of quantum error correction; This means correcting quantum error is not the necessary functionality in quantum public-key cryptosystems; Our scheme removes this redundant functionality, and makes the encoding/decoding more simple.


\section{Conclusions}
Quantum version of McEliece PKE is analyzed and is at least as secure as their classical counterparts, and, at the same time, are also shown that they cannot be more secure than related one-way function. We also suggest double-encryption scheme to improve the security of the QPKE protocol, and analyze its security would decrease while multiply applying the double-encryption scheme.

\section*{Acknowledgements}
This work was supported by the National Natural Science Foundation of China under Grant No. 60573051 and 61173157.


\appendix
\section*{Appendix}
Ref.\cite{Okamoto00} introduces a constant-weight coding algorithm which can encode each $k$-bit messages $m$ to a $n$-bit string $w(m)=e_1e_2\cdots e_n$ of the same weight $t$, and different messages has different codes. This algorithm can be modified to be a quantum encoding alogrithm which implements the transformation \begin{equation}\label{equ5}\sum_m\alpha_m|m\rangle\rightarrow\sum_m\alpha_m|w(m)\rangle.\end{equation}
The number of qubits changes after the above transformation. It is worth to explain why this quantum transformation is valid. Because the encoding algorithm $m\rightarrow w(m)$ is a reversible computing, and both the two-way computing can be implemented efficiently, the following two steps of quantum computing can also be implemented efficiently:
\begin{eqnarray*}
\sum_m\alpha_m|m\rangle|0\rangle &\rightarrow & \sum_m\alpha_m|m\rangle|w(m)\rangle \\
&\rightarrow & |0\rangle\sum_m\alpha_m|w(m)\rangle.
\end{eqnarray*}
Thus, the quantum encoding alogrithm can be written as the Eq.(\ref{equ5}). However, this is not valid for general computation. We prove it in the following propositions 4 and 5.

{\bf Proposition 4:} Quantum transformation $\sum_m\alpha_m|w(m)\rangle\rightarrow\sum_m\alpha_m|w(m)H\rangle$ is infeasible physically, where $H$ is $n\times k$ $(n>k)$ binary matrix, and $w(m)\in\{0,1\}^n$ is a constant-weight code of $m$.

Clearly, the computing $w(m)\rightarrow w(m)H$ can be implemented by a polynomial size classical circuit. However, the reverse computing cannot been finished, because the $n\times k (n>k)$ matrix $H$ does not have right inverse. Thus, we cannot express the quantum transformation $\sum_m\alpha_m|w(m)\rangle\rightarrow\sum_m\alpha_m|w(m)H\rangle$.

Next, we give a strict demonstration.

{\bf Proof:} The computing $w(m)\rightarrow w(m)H,\forall m$, changes $n$-bit string into $k$-bit string. Because $k<n$, it can be think as this:
for arbitrary $n$-bit string $w(m)$, its last $n-k$-bit information is erasured into zeroes and the former $k$ bits is changed to $w(m)H$. This means, there exist a $n\times(n-k)$ binary matrix $A$, such as
\begin{equation}\label{equ7}
w(m)[H|A]=[w(m)H|0\cdots 0], \mathrm{for~all~}n\mathrm{-bit~constant~weight~code}.
\end{equation}
Thus, $w(m)A=0\cdots 0,\forall m$. Then there exists a $n\times(n-k)$ binary matrix $A$ such that, each column of $A$ (denoted as $a_j$) is orthogonal with arbitrary $w(m)$.
Because $w(m)$ is a $n$-bit constant-weight code of weight $t$ ($t<n$), any $t$ elements of any $a_j$ is summed to $0(\mathrm{mod} 2)$.

If $t$ is even, any $a_j$ must be either all-zero vector or all-one vector; If $t$ is odd, $a_j$ must be all-zero vector. Thus, when $t$ is even, each column of $A$ must be either all-zero vector or all-one vector; When $t$ is odd, each column of $A$ must be all-zero vector. No matter which condition happens, the $n\times n$ matrix $[H|A]$ cannot be unitary. So, it is infeasible to physically implement the quantum transformation $\sum_m\alpha_m|w(m)\rangle\rightarrow\sum_m\alpha_m|w(m)H\rangle$.

The result can be extended to the general case.

{\bf Proposition 5:} Quantum transformation $\sum_m\alpha_m|m\rangle\rightarrow\sum_m\alpha_m|mH\rangle$ is infeasible physically, where $H$ is a $n\times k$ $(n>k)$ binary matrix, and $m\in\{0,1\}^n$.

{\bf Proof:} The proof is similar to Proposition 4. In the same way, there exist a $n\times(n-k)$ binary matrix $A$, such as $mA=0\cdots 0,\forall m\in\{0,1\}^n$. Then the matrix $A$ is the all-zero matrix, and the $n\times n$ matrix $[H|A]$ cannot be unitary. So, it is infeasible to physically implement the quantum transformation $\sum\limits_m\alpha_m|m\rangle\rightarrow\sum\limits_m\alpha_m|mH\rangle$. In other words, the transformation $T=\sum_m|\stackrel{n-k}{\overbrace{0\cdots 0}},mH\rangle\langle m|$ is infeasible in physical implementation.

In fact, when part of the amplitudes $\alpha_m,m\in\{0,1\}^n$ are set to zeroes, Proposition 5 would degenerate to Proposition 4.
\end{document}